\documentclass[twocolumn]{article}
\usepackage[english]{babel}
\usepackage{amssymb,amsmath}
\usepackage{subfigure}
\usepackage{graphicx} 
\usepackage{mathrsfs}
\usepackage{comment}
\usepackage{fullpage}
\usepackage{mathrsfs}
\usepackage{bm}
\usepackage{tikz}
\usepackage{amsthm}
\usepackage[sort&compress,numbers]{natbib}
\usepackage[colorlinks,breaklinks,bookmarks,pageanchor,citecolor=blue]{hyperref}

\DeclareFontFamily{U}{wncy}{}
    \DeclareFontShape{U}{wncy}{m}{n}{<->wncyr10}{}
        \DeclareSymbolFont{mcy}{U}{wncy}{m}{n}
            \DeclareMathSymbol{\Sh}{\mathord}{mcy}{"58}

\title{Time-dependent Electronic Populations in Fragment-based Time-dependent Density Functional Theory}

\author{Mart\'in A. Mosquera\textsuperscript{1}\thanks{hhh}, Adam Wasserman\textsuperscript{1,2}\\
\small \textsuperscript{1}Department of Chemistry\\[-0.8ex]
\small Purdue University, 560 Oval Drive, West Lafayette, IN 47907, USA\\
\small \textsuperscript{2}Department of Physics and Astronomy\\[-0.8ex] 
\small Purdue University, 525 Northwestern Avenue, West Lafayette, IN 47907, USA\\
\small \texttt{awasser@purdue.edu}
}

\date{April, 2015}

\begin{document}
\bibliographystyle{unsrtnat}
\newtheorem{thm}{Theorem}
\newtheorem{cor}{Corollary}
\def\bea{\begin{eqnarray}}
\def\eea{\end{eqnarray}}
\def\ben{\begin{equation}}
\def\een{\end{equation}}
\def\benu{\begin{enumerate}}
\def\enu{\end{enumerate}}

\def\n{n}
\def\lsim {\ifmmode {\buildrel<\over\sim}}

\def\sss{\scriptscriptstyle\rm}

\def\g{_\gamma}

\def\t{\beta}
\def\lfc{^{\lambda=1}}
\def\lo{^{\lambda=0}}

\newcommand{\mr}[1]{\mathrm{#1}}
\newcommand{\mc}[1]{\mathcal{#1}}
\newcommand{\ms}[1]{\mathscr{#1}}
\newcommand{\mb}[1]{\mathbf{#1}}
\newcommand{\ud}{\mr{d}}
\newcommand{\ui}{\mr{i}}
\newcommand{\intdr}{\int \ud^3\br~}
\newcommand{\XC}{_{\mr{XC}}}
\newcommand{\HXC}{_{\mr{HXC}}}
\newcommand{\h}{_{\mr{h}}}
\newcommand{\Ha}{_{\mr{H}}}
\newcommand{\s}{_{\mr{s}}}
\newcommand{\up}{_{\mr{p}}}
\newcommand{\dernr}[1]{\frac{\delta {#1} }{\delta n(\br)}}
\newcommand{\dernrs}[1]{\frac{\delta {#1}}{\delta n_{\sigma}(\mb{x})}}
\newcommand{\derN}[1]{\frac{\partial {#1}}{\partial N}}
\newcommand{\dss}{\displaystyle}
\newcommand{\nv}[1]{[n_{v,#1}]}
\newcommand{\snv}[1]{n_{v,#1}}
\newcommand{\KS}{_\mr{KS}}
\newcommand{\LDA}{^\mr{LDA}}
\newcommand{\ELDA}{^\mr{ELDA}}
\newcommand{\rket}[1]{|#1\rangle}
\newcommand{\lket}[1]{\langle #1|}
\def\marnote#1{\marginpar{\tiny #1}}
\def\rsav{\langle r_s \rangle}
\def\invdif{\frac{1}{|\br_1 - \br_2|}}

\def\hatT{{\hat T}}
\def\hatV{{\hat V}}
\def\hatH{{\hat H}}
\def\1var{(\bx_1...\bx\N)}

\def\half{\frac{1}{2}}
\def\quart{\frac{1}{4}}

\def\bp{{\bf p}}
\def\br{{\bf r}}
\def\bR{{\bf R}}
\def\bu{{\bf u}}
\def\b1{{\bf 1}}
\def\bx{{x}}
\def\by{{y}}
\def\ba{{\bf a}}
\def\bq{{\bf q}}
\def\bj{{\bf j}}
\def\bX{{\bf X}}
\def\bF{{\bf F}}
\def\bchi{{\bf \chi}}
\def\bof{{\bf f}}

\def\cA{{\cal A}}
\def\cB{{\cal B}}
\def\cH{{\cal H}}

\def\xj{_{{\sss X},j}}
\def\xcj{_{{\sss XC},j}}
\def\N{_{\sss N}}
\def\H{_{\sss H}}
\def\sH{^{\sss H}}
\def\ext{_{\rm ext}}
\def\pot{^{\rm pot}}
\def\hyb{^{\rm hyb}}
\def\hah{^{1/2\& 1/2}}
\def\LSD{^{\rm LSD}}
\def\TF{^{\rm TF}}
\def\LDA{^{\rm LDA}}
\def\GEA{^{\rm GEA}}
\def\GGA{^{\rm GGA}}
\def\SPL{^{\rm SPL}}
\def\sce{^{\rm SCE}}
\def\PBE{^{\rm PBE}}
\def\DFA{^{\rm DFA}}
\def\VW{^{\rm VW}}
\def\helm{^{\rm unamb}}
\def\una{^{\rm unamb}}
\def\ion{^{\rm ion}}
\def\gs{^{\rm gs}}
\def\dyn{^{\rm dyn}}
\def\adia{^{\rm adia}}
\def\I{^{\rm I}}
\def\pot{^{\rm pot}}
\def\sav{^{\rm sph. av.}}
\def\unif{^{\rm unif}}
\def\LSD{^{\rm LSD}}
\def\ee{_{\rm ee}}
\def\vir{^{\rm vir}}
\def\ALDA{^{\rm ALDA}}
\def\PGG{^{\rm PGG}}
\def\GK{^{\rm GK}}

\def\up{_\uparrow}
\def\dn{_\downarrow}
\def\upp{\uparrow}
\def\dnn{\downarrow}

\def\td{time-dependent~}
\def\KS{Kohn-Sham~}
\def\DFT{density functional theory~}

\def\fourint{ \int_{t_0}^{t_1} \! dt \int \! d^3r\ }
\def\fourintp{ \int_{t_0}^{t_1} \! dt' \int \! d^3r'\ }
\def\intx{\int\!d^4x}
\def\sph_int{ {\int d^3 r}}
\def\radint{ \int_0^\infty dr\ 4\pi r^2\ }

\def\PRA{Phys. Rev. A\ }
\def\PRB{Phys. Rev. B\ }
\def\PRL{Phys. Rev. Letts.\ }
\def\JCP{J. Chem. Phys.\ }
\def\JPCA{J. Phys. Chem. A\ }
\def\IJQC{Int. J. Quant. Chem.\ }

\def\la{{\langle\, }}
\def\ra{{\,\rangle }}
\def\infintw{ \int_{-\infty}^\infty }
\def\infint{ \int_{-\infty}^\infty dx\,}
\def\infintp{ \int_{-\infty}^\infty dx'\,}
\def\infintd3r{ \int_{-\infty}^\infty d^3r\,}
\def\intd3r{ \int d^3r\,}
\def\kinop{- \half \frac{d^2}{dx^2}}
\def\laplace1d{\frac{d^2}{dx^2}}
\def\plaplace1d{\frac{d^2}{d{x'}^2}}
\def\pdr{\frac{\partial}{\partial r}}
\def\padr2{\frac{\partial^2}{\partial r^2}}
\def\nup{n_\uparrow}
\def\ndown{n_\downarrow}
\def\p{\,|\,}
\def\lin{^{(1)}}
\def\bmu{{\vec \mu}}
\def\SMA{^{\rm SMA}}
\def\SPA{^{\rm SPA}}

\def\sF{\boldsymbol{\Phi}}
\def\sP{\boldsymbol{\cal P}}
\def\sV{{\cal V}}
\def\sN{\boldsymbol{\cal N}}
\def\st{\boldsymbol{\tau}}
\def\O{{\cal O}}
\def\F{{\cal F}}
\def\N{{\cal N}}
\def\a{{\alpha}}
\def\b{{\beta}}
\def\P{{\cal P}}
\def\hP{\hat{P}}
\def\tP{\tilde{P}}
\def\E{{\cal E}}
\def\G{{\cal G}}
\def\S{{\cal S}}

\def\bS{{\overline S}}

\twocolumn[
  \begin{@twocolumnfalse}
      \maketitle
\begin{abstract}      
Conceiving a molecule as composed of smaller molecular fragments, or subunits, is  
one of the pillars of the chemical and physical sciences, and leads to productive  
methods in quantum chemistry. Using a fragmentation scheme, efficient algorithms 
can be proposed  
to address problems in the description of chemical bond formation and breaking.  
We present a formally exact time-dependent density-functional theory for the electronic 
dynamics of molecular fragments with variable number of electrons. This new formalism 
is an extension of previous work {[Phys. Rev. Lett. {\bf 111}, 023001 (2013)]}. 
We also introduce a stable density-inversion method that is applicable to  
time-dependent and ground-state density-functional theory, and their extensions, including those 
discussed in this work. 

\end{abstract}
  \end{@twocolumnfalse}
]
{
\renewcommand{\thefootnote}%
  {\fnsymbol{footnote}}
    \footnotetext[1]{Present address: Department of Chemistry, Northwestern University, 2145 Sheridan Road, Evanston, IL 60208,
    USA}
}

\section{Introduction}
Simple and productive methods to investigate dynamical features 
of solids and molecules are offered by Time-dependent Density-functional Theory (TDDFT) \cite{GM12}. 
TDDFT embodies many concepts and formal exact results, but its core 
is the 1-1 correspondence \cite{RG84} between time-dependent (TD) external potentials and TD electronic densities, provided the 
initial state of the system is given. Every observable of the system is expressed as a TD density-functional.
The calculation of the density is carried out by solving the TD Kohn-Sham (KS) equations \cite{P78}, which are 
single-particle Schr\"odinger equations that require an approximation 
to the exchange-correlation (XC) potential, a density-functional. 

The adiabatic local density approximation (ALDA) \cite{ZS80} to the TD XC potential
is the simplest, useful approximation to study the dynamics of atoms and solids. 
However, when applied to molecules, 
ALDA often yields 
unphysical results, for example, atoms with fractional charges at dissociation, underestimated charge-transfer excitation 
energies, missing double excitations, among others.  Two decades of research have shown that it is 
challenging to enhance the performance 
of ALDA while preserving computational simplicity and elegance.   

The TD KS equations describe all of the electrons as part
of a single entity, imposing a limit on the number of atoms that can be simulated 
in a reasonable amount of time. This limit can be increased by dividing a molecule into fragments
and performing inexpensive calculations on each individual fragment. Several approximated methods to 
investigate the electron dynamics of molecules as composed of fragments are available \cite{CW04,CFK07,N10,P13}.
These methods typically assign a set of TD single-particle Schr\"odinger 
equations (not necessarily TD KS equations) to every fragment in the molecule. The fragment electrons are subject to the usual interactions as if 
they were in the presence of the fragment nuclei only; and, an extra potential, usually fragment-dependent, accounting 
for the interaction between the fragments, is added. 
Successful applications to the calculation of solvachromatic shifts \cite{N07,PJ12} and excitation 
energy of monomers \cite{CFK07} have been reported.

A rigorous extension of TDDFT for molecules made of chemical fragments was presented in Ref. \cite{MJW13}. 
In this extension, a molecule is divided into fragments, each a set of atoms. Every fragment is assigned an initial state, and a  
Hamiltonian including a global, auxiliary potential, termed partition potential, which enforces 
that the total electronic density is the true TD electronic density of the molecule. We proved 
that the partition potential is uniquely determined by the TD electronic density of the system, and that  
it can thus be expressed (and approximated) as a density-functional. The linear response and an extension to consider electromagnetic 
fields are presented in Ref. \cite{MW14}. 
The Hamiltonians used in Refs. \cite{MJW13} and \cite{MW14}, and the aforementioned 
approximated methods, are particle-conserving, i.e., 
the average number of electrons in a fragment is time-independent; this restriction is eliminated in this work. 

In this paper we first introduce a density-inversion method that improves that of Ref. \cite{MJW13}
and allows us to estimate the partition potential. Using a model for the errors, we 
discuss, in close connection with standard TDDFT, the   
origin of uncertainties of the partition potential in regions far from the molecule. 
We extend our fragment-based TDDFT \cite{MJW13} to allow for variable numbers of 
electrons in each fragment, while preserving the uniqueness of observables as density-functionals. 
The formalism introduced in this paper serves as a theoretical foundation for the development of 
methods accounting for electronic excitations and 
electron-transfer processes, without sacrificing explicit use of the electronic density and computational efficiency. 

\section{Fragment-based TDDFT}
\subsection{Formulation}
An electron in a fragment, labeled $\alpha$, is subject to a 1-body external potential, 
denoted as $v_{\alpha}$. For example, $v_{\alpha}(\br)=\sum_{i \in I_{\alpha}} -Z_i/|\br-\bR_i|$; 
$I_{\alpha}$ is a set of labels for the nuclei in fragment $\alpha$.
We assign to each fragment in the molecule a Hamiltonian that includes an auxiliary potential, 
the partition potential $v_{\mr{p}}$:
\ben
\hat{H}_{\alpha}[v_{\mr{p}}](t)=\hat{H}_{\alpha}^0+\int \ud^3\mb{r}~ \hat{n}(\mb{r})v_{\mr{p}}(\mb{r}t)~,
\een
where $\hat{H}_{\alpha}^0=\hat{T}+\hat{W}+\int \ud^3\mb{r}~\hat{n}(\mb{r})v_{\alpha}(\mb{r})$, $\hat{T}$ and $\hat{W}$ 
are the kinetic, and coulombic repulsion energy operators, respectively, and $\hat{n}(\br)$ is the density operator. 
This Hamiltonian does not include external driving forces other than those due to 
the nuclei of fragment $\alpha$. 
TD displacement of the positions of the nuclei can be described by introducing a time-dependent Hamiltonian 
where $v_{\alpha}$ is replaced by the corresponding TD fragment-potential, $\sum_{i\in I_{\alpha}} -Z_i/|\br-\bR_i(t)|$.   

The state of a fragment is described by the evolution of the ket $|\psi_{\alpha}[v_{\mr{p}}](t)\rangle$ in Fock space, which 
satisfies the TD Schr\"odinger equation (atomic units are used throughout):
\ben\label{firstevol}
i\partial_t|\psi_{\alpha}[v_{\mr{p}}](t)\rangle=\hat{H}_{\alpha}[v_{\mr{p}}](t)|\psi_{\alpha}[v_{\mr{p}}](t)\rangle~,
\een
where 
\ben
|\psi_{\alpha}(t)\rangle=\sum_M \nu_{\alpha,M}|\psi_{\alpha,M}(t)\rangle~.
\een 
$\{\psi_{\alpha,M}\}$ are kets corresponding to states with integer number of particles
and $\{\nu_{\alpha,M}\}$ are the weight amplitudes of those states.
Kets with different number of electrons are orthogonal, $\langle \psi_{\alpha,M}|\psi_{\alpha,M'}\rangle=0~~,M\neq M'$.
The total density is given as 
\ben
n(\mb{r}t)=\sum_{\alpha} n_{\alpha}(\mb{r}t)~,
\label{density_constraint}
\een
which defines $v_{\mr{p}}$ when 
$
n_{\alpha}(\mb{r}t)=
\langle\psi_{\alpha}(t)|\hat{n}(\mb{r})|\psi_{\alpha}(t)\rangle~,
$
as proven in Ref. \cite{MJW13}: Given a set $\{\psi_{\alpha,0},\,v_{\alpha}\}$, 
two potentials $v_{\mr{p}}$ and $v_{\mr{p}}'$ that differ by more than a TD constant 
{\slshape cannot give rise to the same density}. A corollary of this theorem is that 
there is a TD density-functional that, when evaluated at a given TD electronic density, gives 
the corresponding TD partition potential \footnote{This theorem and corollary were recently used 
in Ref. \cite{HLPC14} to propose an inversion method in the context of embedding potential-functional theory.}. 

The partition potential represents the TD electronic density
of the supermolecule, and is decomposed as follows \cite{MW14}: 
\ben
v_{\mr{p}}(\br t)=v_{\mr{G}}(\br t)+v_{\mr{d}}(\br t)~.
\een
$v_{\mr{G}}$ is a ``gluing potential'', accounting for the correlation between the fragments, and
$v_{\mr{d}}$ is the driving potential the molecule is subject to (e.g. laser field). The gluing potential yields 
the shape of the potential such that the TD electronic density is recovered.
The gluing potential satisfies \cite{MW14}:
\ben
\begin{split}\label{gmotion}
\frac{1}{\ui}\nabla\cdot n(\br t)&\nabla v_{\mr{G}}(\br t)=\langle\psi(t)|[\hat{H}^0,\nabla\cdot\hat{\mb{j}}(\br)]|
\psi(t)\rangle\\
        &-\sum_{\alpha}\langle\psi_{\alpha}(t)|[\hat{H}^0_{\alpha},\nabla\cdot\hat{\mb{j}}(\br)]|\psi_{\alpha}(t)\rangle~.
\end{split}
\een
The terms on the right-hand-side of Eq.(\ref{gmotion}) are TD density-functionals. 
Approximating these terms and solving the 
resulting differential equation yields an estimate of the gluing potential. Another 
route to approximating $v_{\mr{G}}$ is 
by assuming that the system evolves adiabatically from its ground states, driven 
by a very slowly-varying field. In such case, the potential $v_{\mr{G}}$ is obtained from 
the adiabatic approximation in ground-state Partition DFT \cite{EBCW10,MW12}: 
\ben
v_{\mr{G}}^{\mr{Ad}}[n(t)]=v_{\mr{p}}^{\mr{Ad}}[n(t)]-v^{\mr{HK}}[n(t)]~,
\een
where $v^{\mr{HK}}[n(t)]$ is the external perturbation the interacting electrons are subject to in their 
ground-state in order to yield the density $n(\mb{r}t)$ (the uniqueness of $v^{\mr{HK}}$ follows from the Hohenberg-Kohn theorem). 
The partition potential, $v_{\mr{p}}^{\mr{Ad}}[n(t)]$, is the Lagrange multiplier required 
to solve the minimization:
\ben
\min_{\{\psi_{\alpha}\}\rightarrow n(t)} \sum_{\alpha}\lket{\psi_{\alpha}}\hat{H}_{\alpha}^0\rket{\psi_{\alpha}}~.
\een
under the constraint of Eq.(\ref{density_constraint}).
The Lagrange multiplier for this problem is unique, up to an arbitrary constant \cite{CW06}.

The TD partition KS equations are:
\ben
\begin{split}
\ui\partial_t &\phi_{i,\alpha}(\br,t)=\Big(-\frac{1}{2}\nabla^2+v_{\mr{Hxc}}[n_{\alpha}](\br,t)
\\ &
+v_{\alpha}(\br)+v_{\mr{G}}[n](\br,t)+
        v_{\mr{d}}(\br,t)\Big)\phi_{i,\alpha}(\br,t)~.
\end{split}
\een
The density can be obtained by means of:
$
n(\br,t)=\sum_{i\alpha} f_{i\alpha}|\phi_{i\alpha}(\br,t)|^2~,
$
where
$\{f_{i\alpha}\}$ are the (time-independent) occupation numbers, chosen from a proper ensemble \cite{MJW13}. 

\section{Classical Interpretation of the Partition Potential}
Consider a system composed of a single massive particle, and a bath 
made of particles much smaller than the massive one. All the particles 
in the particle+bath system interact via a potential, $U_{\mr{int}}$, of the form 
$\sum_{i>j} u_{ij}(\mb{r}_i-\mb{r}_j)$.
The evolution of the subsystem particle, labeled $\mr{S}$, 
is dictated by Eq. (\ref{firstevol}). There is no partitioning 
of the external potential because the particle and the bath are subject, implicitly, to a 
macroscopic, external, confining potential. The Hamiltonian of the subsystem particle is 
thus $\hat{H}_{\mr{S}}=\hat{T}+\int \ud^3\mb{r}~\hat{n}(\mb{r})v_{\mr{p}}(\mb{r}t)$, and the 
Hamiltonian of the bath is $\hat{H}_{\mr{B}}=\hat{T}+\hat{W}+\int \ud^3\mb{r}~\hat{n}(\mb{r})v_{\mr{p}}(\mb{r}t)$.
The average position of the particle 
is $\bar{\br}_{\mr{S}}(t)=\int \ud^3\br~\br~ n_{\mr{S}}(\mb{r},t)$, where
$n_{\mr{S}}(\mb{r},t)=|\psi(\mb{r},t)|^2$. 

By the Ehrenfest theorem and 
correspondence principle we have
\ben
m_{\mr{S}}\frac{\ud^2 \bar{\br}_{\mr{S}}}{\ud t^2}=\mb{F}_{\mr{p},\mr{S}}(t)~,
\een
where $\mb{F}_{\mr{p},\mr{S}}(t)=-\int \ud^3\mb{r}~n_{\mr{S}}(\br,t)\nabla v_{\mr{p}}(\mb{r},t)$, 
and $m_{\mr{S}}$ is the mass of the particle. 
Comparison with the equation of motion of the real system indicates that, in the classical limit, 
$(\nabla v_{\mr{p}})(\bar{r}_{\mr{S}}(t))=(\nabla_{\mb{r}_{\mr{S}}} U_{\mr{int}})
(\bar{r}_{\mr{S}}(t),\bar{\mb{r}}_{\mr{B}}(t))$, where $\bar{\mb{r}}_{\mr{B}}$ is the average coordinate of the bath. 
The partition force at the position of the particle is simply the force exerted on the particle by the bath. 

As the mass of the subsystem particle is increased, the density tends to a Dirac 
distribution. It follows from the above result and Eq. (\ref{gmotion}) that the shape of 
the partition potential for any point but that of the particles
is indefinite. However, for given initial momenta and coordinates of the particles and bath 
(this condition replaces the requirement that the initial wavefunction be specified), the trajectory 
of the momenta of the total system is in a one-to-one correspondence with the trajectory of partition 
forces exerted on each particle. Furthermore, if the assumptions of Langevin dynamics are applicable, the partition force on 
the massive particle can be interpreted as $\mb{F}_{\mr{p},\mr{S}}(t)=-\gamma \mb{v}_{\mr{S}}(t)+\mb{F}_{\mr{ran}}(t)$, 
where $\mb{v}_{\mr{S}}(t)=\ud \bar{\br}_{\mr{S}}/\ud t$, $\gamma$ is the friction coefficient, and $\mb{F}_{\mr{ran}}$
is the random force. 

The gradient of $v_{\mr{p}}$ can {\slshape only} be known at the position of the particles, 
nowhere else. In the original proof by Runge and Gross \cite{RG84}, and its extension for 
partition potentials \cite{MJW13}, it was found that 
potentials whose gradients grow rapidly in regions distant from the molecule 
are not covered. This result, as illustrated in Section 4, is related 
to the uncertainty in the estimation of the partition potential near the boundaries of the 
simulation box.

\section{Numerial TD Potentials}
TDDFT, in which our formulation is built upon, concerns itself with the simplification of the problem:
\ben
(\ui\partial_t -\hat{H}^{\lambda}[v](t))|\psi(t)\rangle=0,\quad \rket{\psi(0)}=\rket{\psi_0}~,
\een
where
\ben
\hat{H}^{\lambda}[v](t)=\hat{T}+\lambda \hat{W}+\int \ud^3 \mb{r}~ \hat{n}(\mb{r})v(\mb{r}t)~.
\een
Runge and Gross \cite{RG84} showed that if $v$ is Taylor-expandable around $t=0$ and does not have physical 
anomalies in the boundaries, then $v$ determines $n$ uniquely up to a TD constant in the potential;
this theorem (recently used to formulate quantum control in TDDFT \cite{CWG12}) can be extended to include non-analytic potentials \cite{RvL11}. 
Let us denote the RG map as $\Lambda_{\psi_0}^{\lambda}$; thus, $n(t)=\Lambda_{\psi_0}^{\lambda}[v](t)$. 
The operator $\hat{W}$ can stand for different types of electron-electron interactions, such 
as screened coulombic repulsion. If $\lambda=0$, then the electrons are non-interacting.

Suppose a well behaved density, $n^{\mr{ref}}$, and an initial state, $\psi_0$, are known. If 
$v_1$ and $v_0$ exist, where $v_{\lambda}(t)=(\Lambda_{\psi_0}^{\lambda})^{-1}[n^{\mr{ref}}](t)$, 
then the Hartree-exchange-correlation potential for the system, by definition, is $v_{\mr{HXC}}=v_0-v_1$. 
Using the exact map $\Lambda_{\psi_0}$ would require solving the TD Schr\"odinger equation, which is what one wants to  
avoid in practical calculations. 

For the development of functionals, exploration of the map
$\Lambda_{\psi_0}^{\lambda}$ is fruitful; this map could be investigated by solving the problem 
$n^{\mr{ref}}(t)-\Lambda_{\psi_0}^{\lambda}[v](t)=0$, which is a root-finding problem. 
The first order response of the density 
for some perturbation $\delta v$ is $\delta n(\mb{r}t)=\int \ud^3\mb{r}\ud t'~\chi^{-1}(\mb{r}t,\mb{r}'t')
\delta v(\mb{r}'t')$. The response function $\chi^{-1}$ should decay in the asymptotic regions because 
large perturbations of $v$ in those regions have a small response in $n$. This relation between 
densities and potentials introduces instabilities in root-finding algorithms aimed at reproducing the density 
corresponding to a given external potential. 
In the ground-state case, the instabilities could be eliminated by enforcing satisfaction of eigenvalue 
constraints. For three dimensional applications, in general, capturing the asymptotic regions is difficult when 
Gaussian basis sets are used because they do not have the correct asymptotic behavior. 

Instead of looking for the exact root, one can solve a minimization problem:
\ben
\min_{v\in \mc{V}}\int_0^T \lVert n^{\mr{ref}}(s)-\Lambda^{\lambda}_{\psi_0}[v](s)\rVert^2~ \ud s~,
\een
where $\lVert f \rVert$ is a suitable spatial norm, and $\mc{V}$ is a space of physical potentials. 
The quantities $n^{\mr{ref}}(t)$ and 
$\langle \psi[v](t)|\hat{n}(\mb{r})|\psi[v](t)\rangle$ need to be approximated. Let us write 
$n^{\mr{ref}}(t)-\Lambda_{\psi_0}^{\lambda}[v](t)=\tilde{n}^{\mr{ref}}(t)-\tilde{\Lambda}_{\psi_0}^{\lambda}[v](t)+
\epsilon[n^{\mr{ref}},v]$. $\tilde{n}^{\mr{ref}}(t)$ is the approximation to $n^{\mr{ref}}(t)$ and 
$\tilde{\Lambda}_{\psi_0}^{\lambda}[v]$ is the approximation to $\Lambda_{\psi_0}^{\lambda}$. 
If $v^*$ is the exact potential representing $n^{\mr{ref}}$, then the problem becomes $\tilde{n}^{\mr{ref}}=
\tilde{\Lambda}_{\psi_0}^{\lambda}[v^*]+\epsilon$. Because we cannot use exact methods to determine 
$n^{\mr{ref}}$ and $\Lambda_{\psi_0}^{\lambda}$, we assume that $\epsilon$ is a random function of the space-time
coordinates. Moreover, one would 
expect that $\tilde{n}^{\mr{ref}}$ and $\Lambda_{\psi_0}^{\lambda}$ have smooth timespace gradients, and that 
$\epsilon$ displays autocorrelation because the spacing between points is arbitrarily small. 

\subsection{Estimation of the Partition Potential}
Let $\mc{V}_{\mr{p}}$ be a space of TD partition potentials, and $\mc{D}$ a space of TD densities and define 
the map:
\ben
\Lambda_{S_0}:\mc{V}_{\mr{p}}\rightarrow\mc{D}~,
\een
where $S_0=\{\psi_{\alpha,0},v_{\alpha}\}$. For a given TD partition potential, the density 
is obtained by evaluation of the above map at the given partition potential; in other words, 
$n(t)=\Lambda_{S_0}[v_{\mr{p}}](t)$. This map depends on the history of the partition potential, 
i.e., it has memory dependence \cite{MJW13}.  

Let $v_{\mr{p}}^*$ be the true partition potential. We assume that, due to numerical 
errors, the estimation to the reference density $\tilde{n}^{\mr{ref}}$ is of the 
form $\tilde{n}^{\mr{ref}}=\tilde{\Lambda}_{S_0}+\epsilon$, where $\epsilon$
is a {\slshape random function}.
To estimate the partition potential corresponding to $\tilde{n}^{\mr{ref}}$ we minimize:
\ben
\lVert e[v_{\mr{p}}]\rVert^2_{\mu}=\lVert \tilde{n}^{\mr{ref}}-\tilde{\Lambda}_{S_0}[v_{\mr{p}}]\rVert^2_{\mu}~,
\een
where $\ud\mu(\mb{r},t)$ is the measure. 

Since $\epsilon$ is a function, its probability density function (PDF) is a {\slshape functional}. 
The PDF depends on parameters, denoted collectively as $\Theta$, and the PDF itself as $D([\epsilon]|\Theta)$.
The probability that $\epsilon$ is observed in a set $\mc{U}$ is given by the path integral:
\ben
P(\epsilon\in \mc{U}|\Theta)=\int_{\mc{U}}\ud m_{\mr{L}}[\epsilon]~ D([\epsilon]|\Theta)~,
\een
where the measure over the space of errors is $m_{\mr{L}}$. The traditional methods 
of non-linear regression can be applied to estimate the best parameters of the distribution, $\Theta^*$, 
for a given set of observations. Then a Taylor expansion in terms of the parameters can be used 
to generate their PDF, which can then be used to estimate the error in the parameters. 
In this case, the parameters are the variance and the partition potential.

In the next section, we will expand the partition potential in a spline basis set.   
The parameters are the values of the partition potential at the knots, and they  
are correlated: A perturbation of the partition potential at 
one knot affects the response of the density in other knots. Hence, we must employ 
a model of correlated errors. Finding the correct model is a demanding task
beyond the scope of this work. For this reason, we choose a biased model based 
on the following observations: i) A measure of the error of the form 
$ 
\int \ud^3\br\ud t~(\tilde{n}^{\mr{ref}}(\br,t)-\tilde{\Lambda}_{S_0}[v_{\mr{p}}])^2
$
suffers of autocorrelation. ii) Far from the molecule, the partition potential 
has little influence on the density. iii) Estimating the density is not sufficient; 
its spatio-temporal gradient is an important quantity. An error measure accounting 
for these observations is:
\ben\label{err_fun}
\lVert e[v_{\mr{p}}]\rVert^2_{\mu}=
\int \ud\mu(\br,t)\{ |\nabla e(\br,t)|^2+(\partial_t e(\br,t))^2\}~.
\een
Based on ii), we choose a measure of the form $\ud\mu(\br,t)=\ud^3\br\ud t~\sum_i \tilde{n}^{\mr{ref}}(\br_i,t)\delta(\br-\br_i)$.
Where $\{\br_i\}$ are points selected in such a way that $|\nabla e|^2+(\partial_t e)^2$ resembles
a $\chi^2$-distribution. 
To apply this measure of error in the next section, we transform 
the above measure into a vector norm. 
Then, the resultant distribution is expanded 
in terms of the partition potential evaluated at the knots, and asymptotic analysis 
is applied \cite{SW89}, leading to the random variables required to reproduce the density 
within a small error tolerance.

\subsection{1d Electron in a Double-well Potential}\label{texample}
Let us revisit the example considered in Ref. \cite{MJW13}: 
A one dimensional ``electron" in a double well potential.
The TD partition equations are: 
\ben
\ui\partial_t \phi_{\alpha}(x,t)=\Big(-\frac{1}{2}\partial^2_x+v_{\alpha}(x)+v_{\mr{p}}(x,t)\Big)\phi_{\alpha}(x,t)~,
\een
where $\alpha=\mr{L},~\mr{R}$, standing for left and right wells. 
The potentials are $v_{\alpha}(x)=v_0/\sqrt{(x-x_{\alpha})^2+a}$; the 
parameters are: $v_0=-1$, $x_{\mr{R}}-x_{\mr{L}}=4$, and $a=1$.
The density is obtained by averaging over the orbital densities of both wells:
\ben
n(x,t)=\frac{1}{2}|\phi_{\mr{L}}(x,t)|^2+\frac{1}{2}|\phi_{\mr{R}}(x,t)|^2~.
\een
Suppose that the supermolecule evolves from the ground state driven by a monochromatic laser. The 
evolution of the system is thus dictated by the solution of:
\ben
\ui\partial_t\psi(x,t)=\Big(-\frac{1}{2}\partial^2_x+v(x)+v_{\mr{d}}(x,t)\Big)\psi(x,t)~, 
\een
where $v_{\mr{d}}(xt)=Ex\sin \omega t$, and the external potential is $v=v_{\mr{L}}+v_{\mr{R}}$.
The density obtained from the above evolution equation is $n^{\mr{ref}}(xt)=|\psi(xt)|^2$, 
which is the target density we wish to represent.
\begin{figure}[tbh!]
\centering
\subfigure[$v$ and $v_{\mr{p}}^0$]{
\includegraphics[scale=0.20]{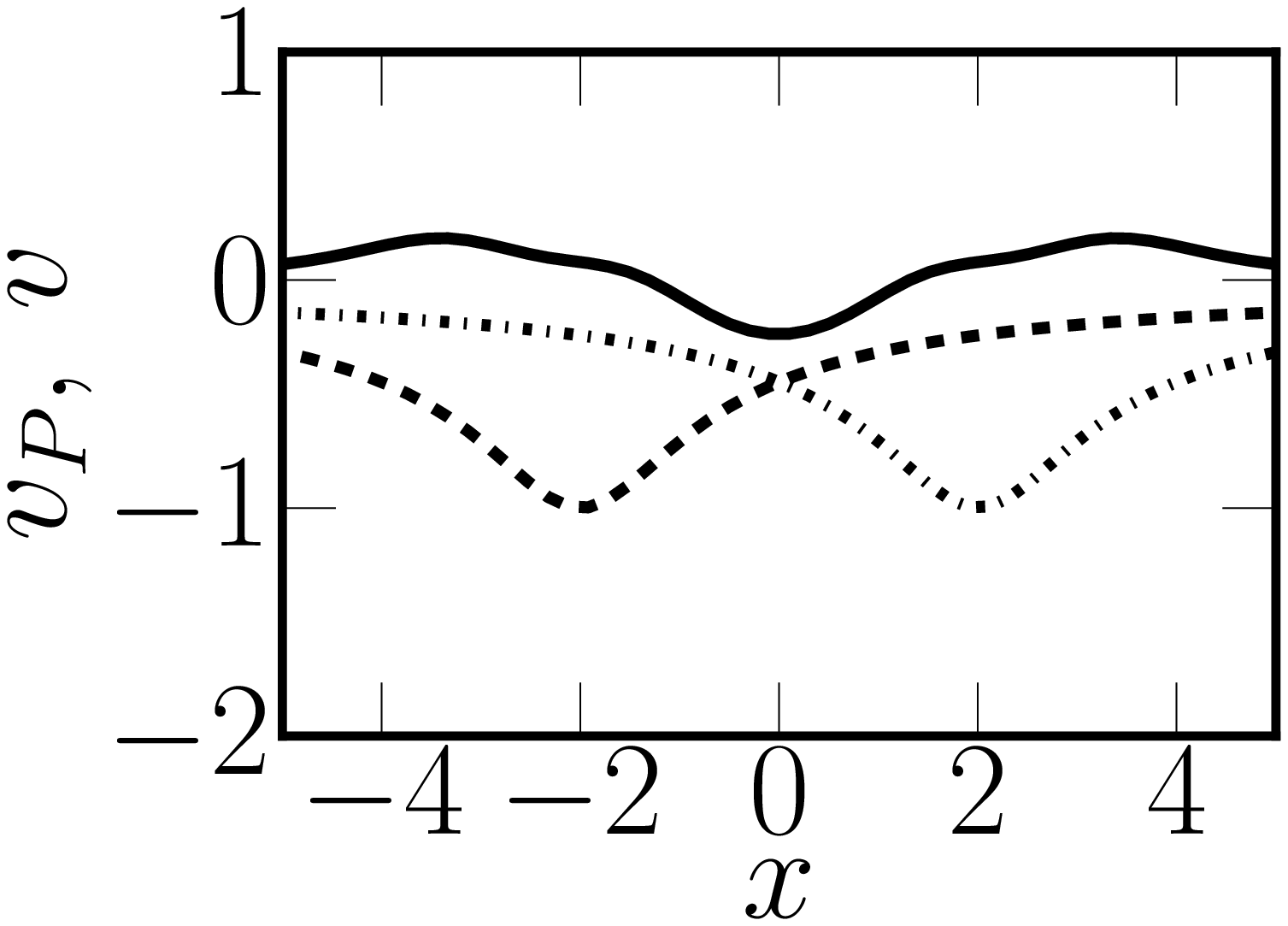}}
\subfigure[Initial Densities]{
\includegraphics[scale=0.20]{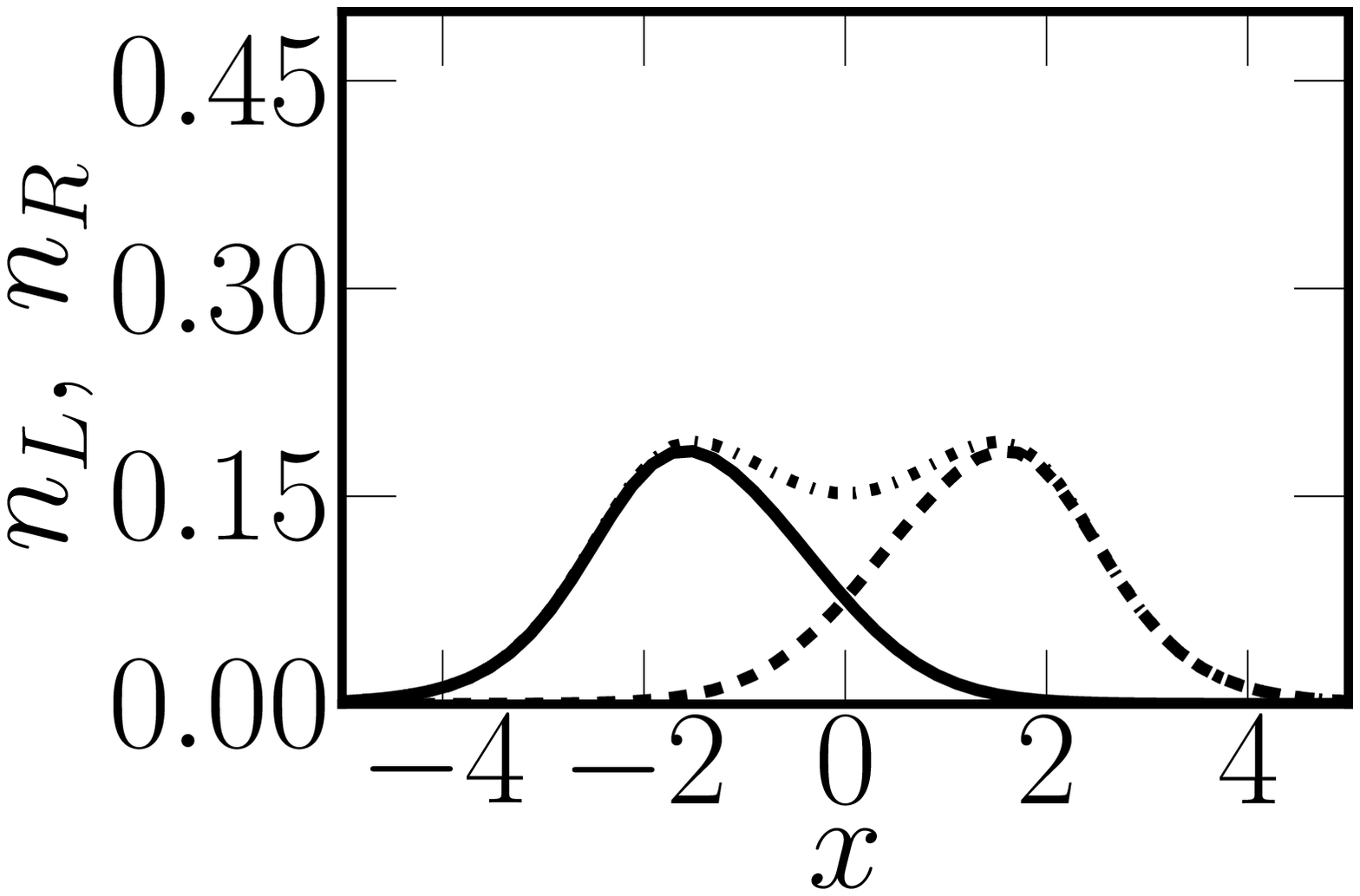}}
\\
\subfigure[$v_{\mr{p}}(t=1.0)$]{
\includegraphics[scale=0.20]{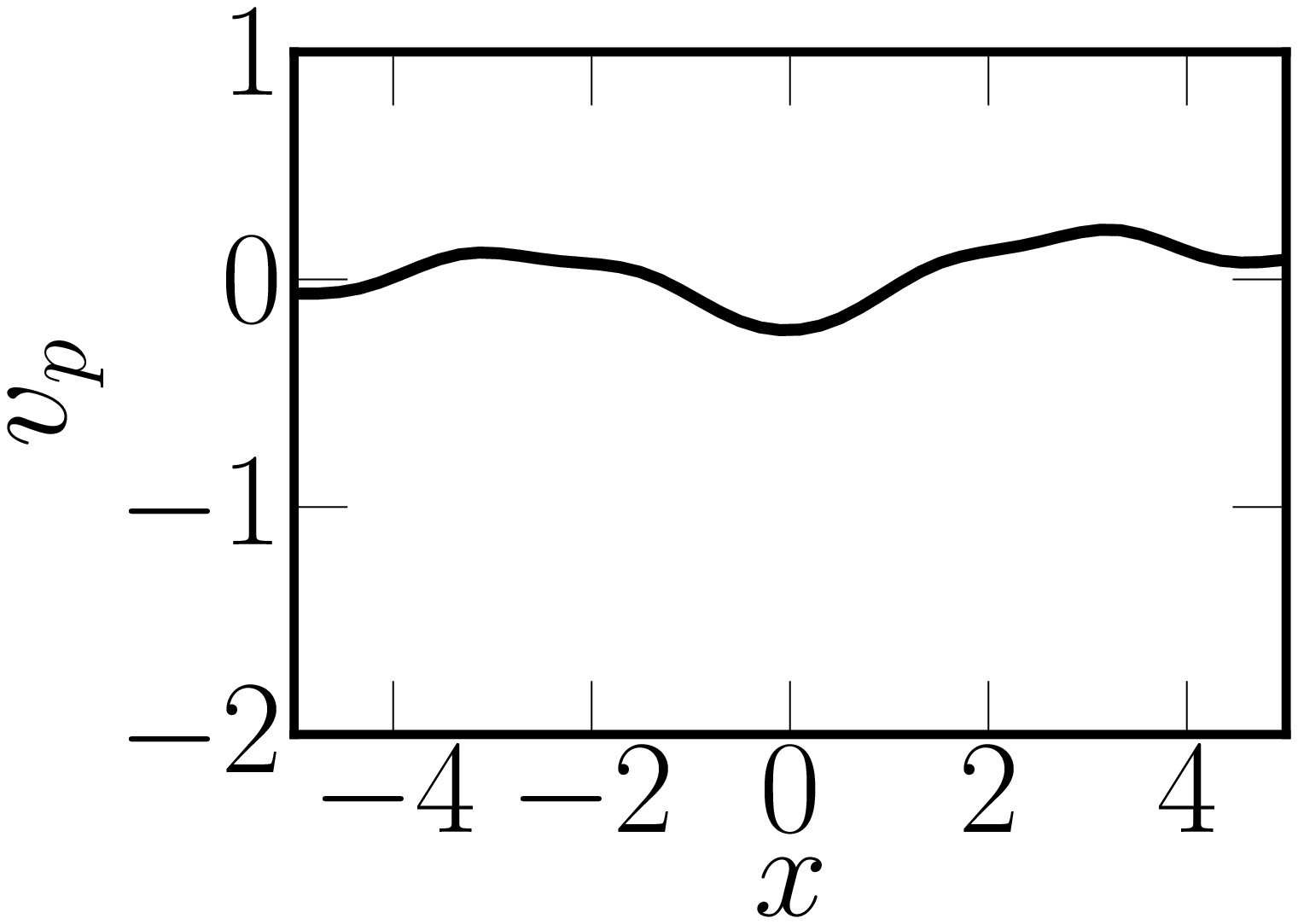}}
\subfigure[$n(t=1.0)$]{
\includegraphics[scale=0.20]{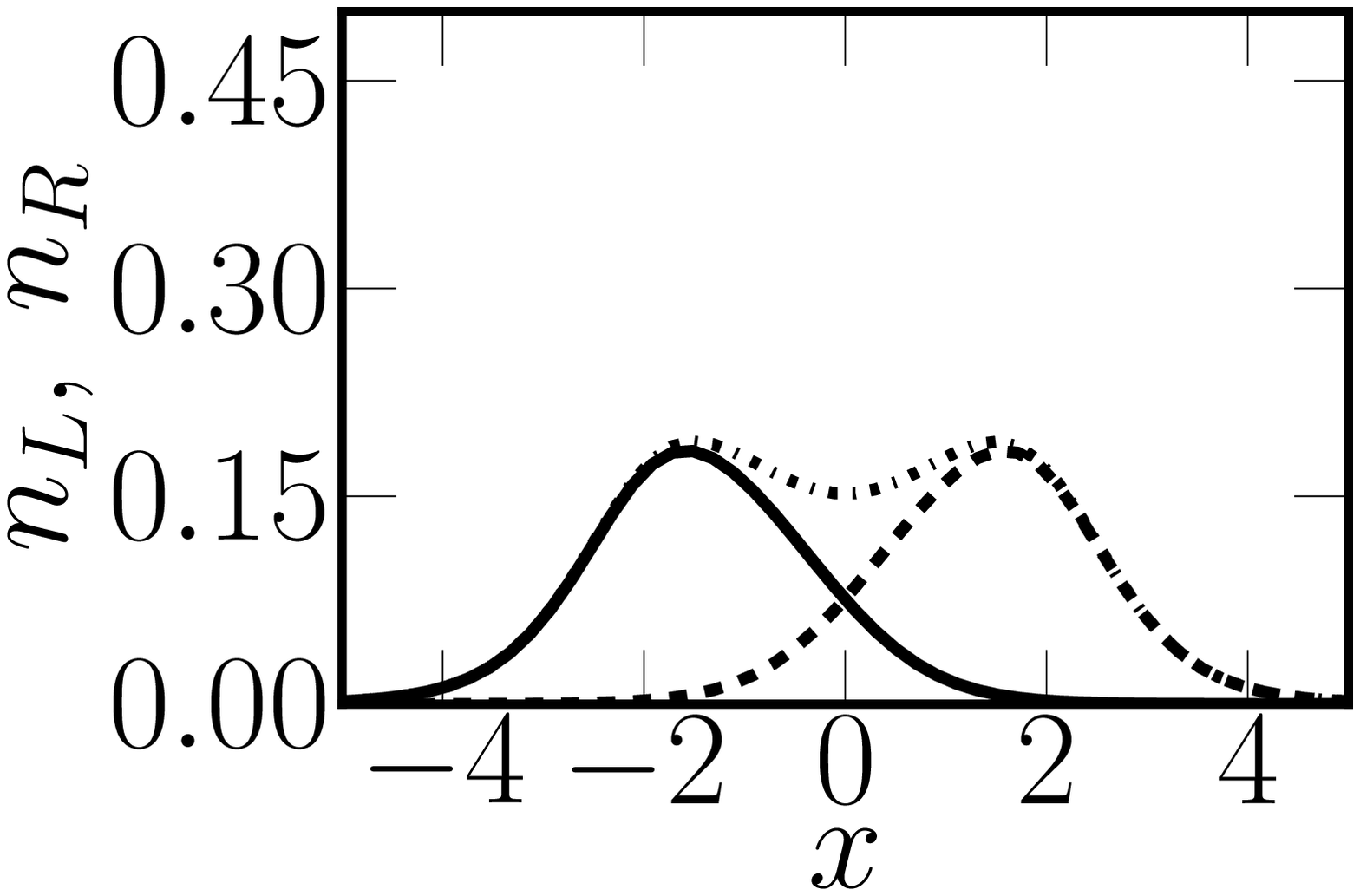}}\\
\subfigure[$v_{\mr{p}}(t=8.0)$]{
\includegraphics[scale=0.20]{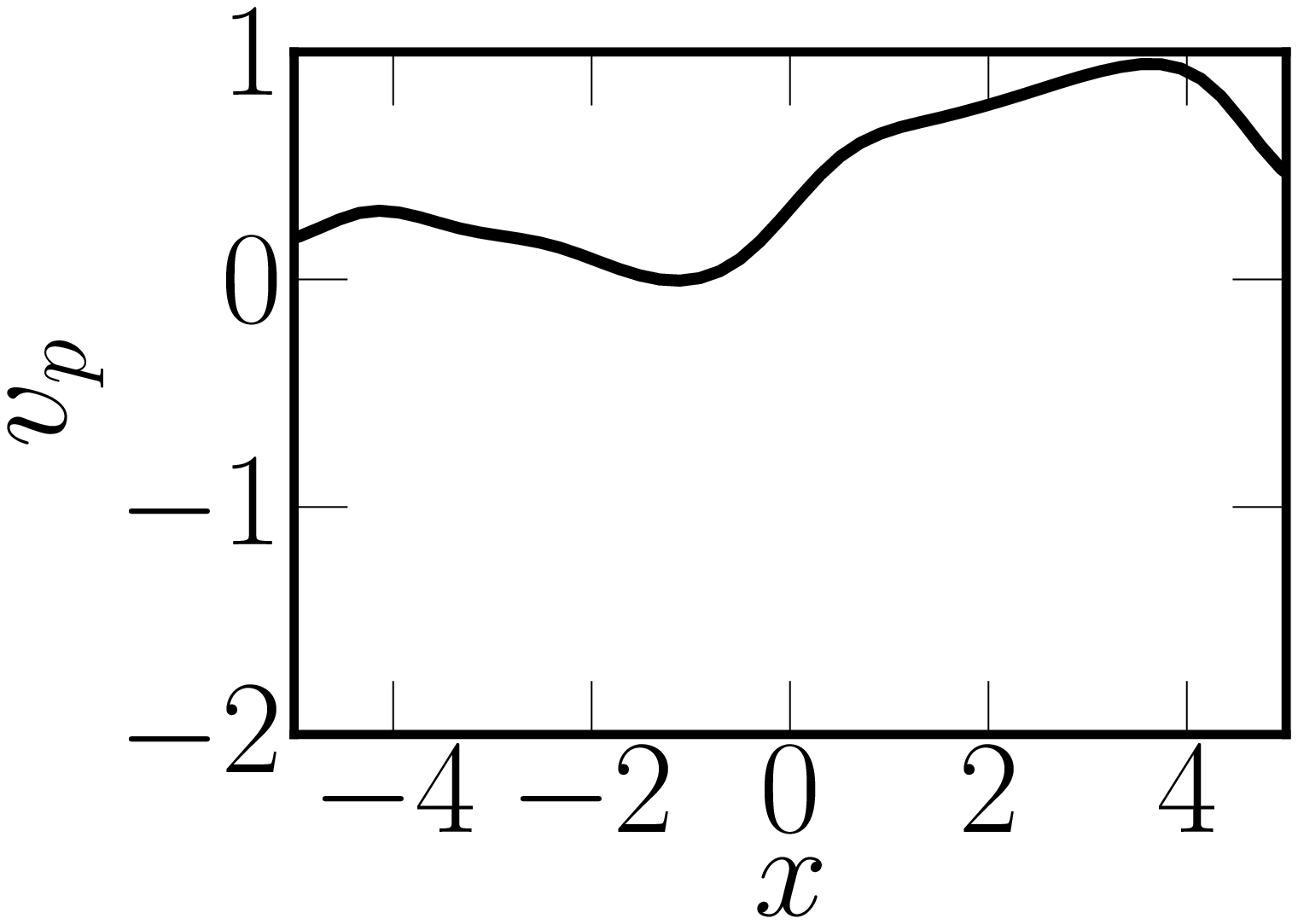}}
\subfigure[$n(t=8.0)$]{
\includegraphics[scale=0.20]{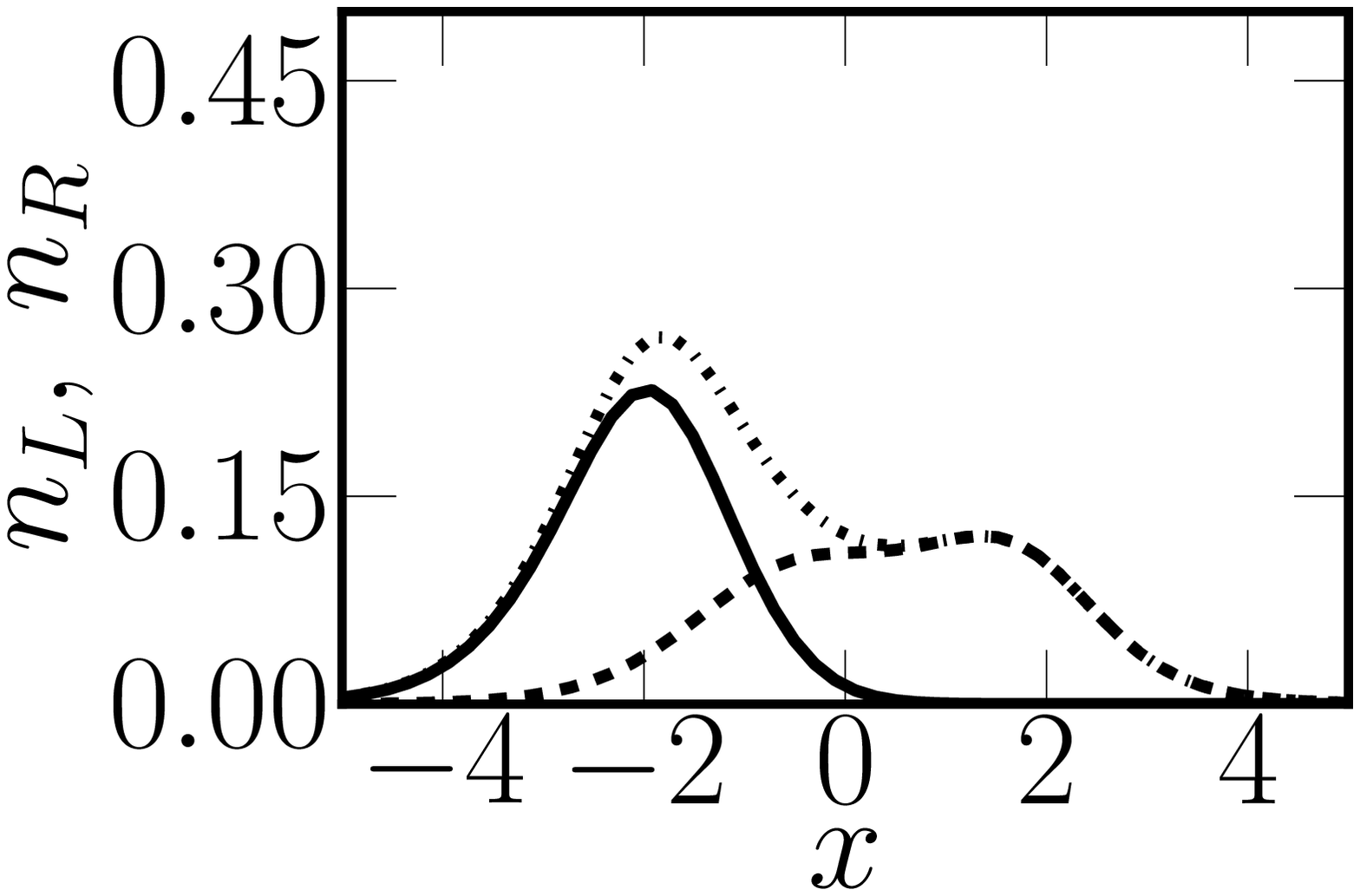}}
\caption{Snapshots of the partition potential. In a), solid line: Initial partition potential, 
dashed line: Left fragment external potential, dashed-dotted line: Right fragment external potential. 
In b), d), and f), solid lines: Left electronic density, dashed lines: Right electronic 
density. In d) and f) the dashed-dotted line is the total density.}
\end{figure}

The laser parameters are $\omega=0.3$, $E_0=0.05$.
We propagate the states of the system using the Crank-Nicholson method;
time step is 0.1, box length is 20, spatial step is 0.2, and total 
propagation time is 10 units.
The partition potential is represented in a smoothed, cubic, spline basis set with 22 knots equally spaced in the box. The 
initial partition potential is estimated by minimizing the error using sequential quadratic programming (other 
useful methods employ the density error directly \cite{RABB15}). 
To obtain the initial densities, the error functional of Eq. (\ref{err_fun}) (with the time-dependency dropped) 
is minimized.
First the problem $(-1/2\partial_x^2 +v_{\alpha}+v_{\mr{p}}^0)\phi_{n,\alpha}=\epsilon_{n,\alpha}\phi_{n,\alpha}$
is solved (with the finite-difference method) for both wells with some estimation of $v_{\mr{p}}^0$. Then, 
the density is compared with that of the system of reference in order 
to propose the next estimation in the iterative procedure of sequential quadratic programming; 
the constraints are conservation of charge and chemical potential (HOMO) equalization \cite{MW14}.
Figure 1.a. shows the initial partition potential and external potentials of each well. 
The initial fragment densities that add up to the ground-state density of the supermolecule 
are displayed in Figure 1.b. 

The estimation of the evolution of $v_{\mr{p}}$ is determined by using the 
step-by-step scheme proposed in Ref. \cite{MJW13}, and the norm of Eq. (\ref{err_fun}):  
For example, for $\Delta t=0.1$, the error norm can be approximated as 
\ben
\Delta t \sum_i \{\tilde{n}^{\mr{ref}}(x_i,\Delta t)[(\partial_x e(x_i,\Delta t))^2+(\partial_t e(x_i,\Delta t))^2]\}~.
\een
The numerical value of the above quantity depends on the value of $v_p$ at the spline knots 
and at $t=\Delta t/2$. Hence, this quantity is varied until the above function is minimized and 
the total density of the fragments match that of the true system. The procedure 
is repeated similarly for the rest of the propagation.
Figure 1.c. shows the partition potential at $t=1.0$;
it is localized in the intermediate region between the fragments. 
The fragment electronic densities (Figure 1.d) are also well localized at $t=1.0$.
Because in absence of the partition potential the fragment-densities would just be localized
around their wells, the partition potential must be such that it induces the transfer 
of charge from the right fragment into the left fragment (Figure 1.e). However, as we note in Figure 1.f,
the charge transfer in this case is represented by the spreading of the right electronic density 
into the left one, not by a change in the fragment populations. Two observations: i) If one were to assign a grid 
that is fine enough around the center of the wells and then coarser as one moves away from the wells, 
then to describe the density spreading, the grid would need to be time-dependent.
ii) The partition potential must induce the charge transfer and act like a ``spoon''.  

\begin{figure}[tbh!]
\centering
\includegraphics[scale=0.35]{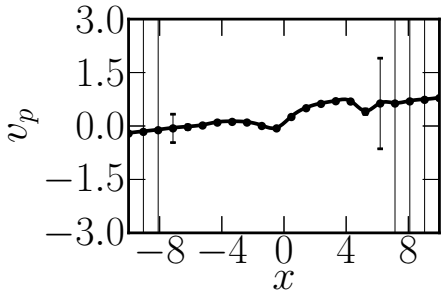}
\caption{Error-bar plot of the partition potential at $t=6.2$. Solid line: 
Interpolated optimal potential, circles: Interpolation knots.}
\end{figure}

The result of the error estimation in the partition potential at $t=6.2$ is shown in 
Figure 2. As expected, in the boundary regions, the error is quite significant, and the error bars extend well beyond the plotting range. 
This implies that the shape of the potential 
in these regions is not reliable. All space-time points obeying causality are coupled. 
For example, variation of a single knot in the boundary affects its neighbors, 
introducing large gradient fluctuations. 
Thus, the error can spread to regions where the density is non-negligible.
This can cause instabilities in the minimization procedure if a norm such as 
$\int \ud^3\mb{r}\ud t~ (\tilde{n}^{\mr{ref}}(\mb{r},t)-n(\mb{r},t))^2$ is employed. For this reason 
we recommend the use of derivatives of the density to measure the error. 

\section{Variable Occupation Numbers}
Let us assign variable electron-occupation numbers to the fragments. 
First, divide the total propagation time into blocks:$[0,\tau)\cup [\tau,2\tau)\cup\ldots\cup [(m-1)\tau,m\tau)$, 
where $m\tau$ is the final time of the propagation, and let 
\ben
X_{\alpha}=\{|\xi_{\alpha}^0\rangle,|\xi_{\alpha}^1\rangle,\ldots,|\xi_{\alpha}^{m-1}\rangle\}~,
\een
be a set of instantaneous kets, in Fock space, for fragment $\alpha$.  
At a single time $t=k\tau$, the following minimization is performed to obtain
the set of kets describing the density of the fragmented molecule:
\ben\label{pdftmin}
\begin{split}
\{|\xi_{\alpha}^k\rangle\}_{\alpha=1}^{N_{\mr{frag}}}=
\arg &\min\Big\{\sum_{\alpha}\langle\psi_{\alpha}^k|
\hat{H}_{\alpha}^0|\psi_{\alpha}^k\rangle ~~\mr{s.t.}\\& \sum_{\alpha}\langle \psi_{\alpha}^k|\hat{n}(\mb{r})|
\psi_{\alpha}^k\rangle= n(\mb{r},k\tau),~\forall \mb{r}\Big\}~,
\end{split}
\een
The occupation numbers of fragment $\alpha$ are formally expressed as 
$ |\nu_{\alpha,M}(k\tau)|^2=|\langle \xi^k_{\alpha,M}|\xi^k_{\alpha}\rangle|^2~.$
Here, $|\xi^k_{\alpha,M}\rangle$ is an optimal ket (obtained from solving the right hand side of Eq. (\ref{pdftmin})) 
for fragment $\alpha$ with $M$ electrons. 
These numbers are density-functionals. 

The evolution operator of fragment $\alpha$ is:
$
\hat{U}_{\alpha}[v_{\mr{p}}](t_1,t_2)=\mc{T}\exp(-\ui\int_{t_0}^{t_1}\ud s~\hat{H}_{\alpha}[v_{\mr{p}}](s))~.
$
Introduce the displaced set of kets:
\ben
\begin{split}
\tilde{X}_{\alpha}&=\{\hat{U}_{\alpha}(\tau,0)|\xi_{\alpha}^0\rangle,\\&\hat{U}_{\alpha}(2\tau,\tau)|\xi_{\alpha}^1\rangle,\ldots
    \hat{U}_{\alpha}(m\tau,(m-1)\tau)|\xi_{\alpha}^{m-1}\rangle\}~.
\end{split}
\een
Now let us define the following dyadic product: $(X_{\alpha}\tilde{X}^{\dagger}_{\alpha})(k)=|\xi_{\alpha}^k\rangle\langle \tilde{\xi}_{\alpha}^{k-1}|$.
The symbol $X_{\alpha}\tilde{X}_{\alpha}^{\dagger}$ is the set of dyadic products where the $k$-th component 
is the dyadic product between the ket at the beginning of the $k$-th block and the 
displaced ket from the $k-1$-th block. 
Now, let $\hat{B}_{\alpha}$ be the TD operator:
\ben
\begin{split}
\hat{B}_{\alpha}(t)&=(W_{\tau}*\ln X_{\alpha}\tilde{X}^{\dagger}_{\alpha})(t)\\
&=\sum_{k=1}^{m}\delta(t-k\tau)\ln |\xi_{\alpha}^k\rangle\langle \tilde{\xi}_{\alpha}^{k-1}|~.
\end{split}
\een
where $W_{\tau}$ is the Dirac-Comb kernel. Addition of the operator $\hat{B}_{\alpha}$ to the Hamiltonian 
$\hat{H}_{\alpha}(t)$ yields the non-Hermitian operator:
\ben
\hat{H}_{\mr{c},\alpha}[v_{\mr{p}}](t)=\hat{H}_{\alpha}[v_{\mr{p}}](t)+\ui\hat{B}_{\alpha}[v_{\mr{p}}](t)~.
\een
The evolution of the system is now determined by $|\psi_{\alpha}[v_{\mr{p}}]\rangle$, which obeys 
\ben
i\partial_t|\psi_{\alpha}[v_{\mr{p}}](t)\rangle=\hat{H}_{\mr{c},\alpha}[v_{\mr{p}}](t)|\psi_{\alpha}[v_{\mr{p}}](t)\rangle~.
\een
The total density is $n(\mb{r}t)=\sum_{\alpha}\langle \psi_{\alpha}(t)|\hat{n}(\mb{r})|\psi_{\alpha}(t)\rangle$ and the 
number of particles in fragment $\alpha$ is $N_{\alpha}(t)=\langle \psi_{\alpha}(t)|\hat{N}|\psi_{\alpha}(t)\rangle$.
In general, any observable, $\hat{O}(t)$, is expressed as a functional of the partition potential, 
$\langle \psi_{\alpha}[v_{\mr{p}}](t)|\hat{O}(t)|\psi_{\alpha}[v_{\mr{p}}]\rangle$.  

Given the partition potential and occupation numbers as density-functionals, the 
scheme to determine the evolution of the molecule is the following:
First, propagate the kets $\{\rket{\psi_{\alpha}}\}$ in the interval $[0,\tau)$ with fixed populations on each 
fragment. Then, at $t=\tau$, obtain new occupation numbers according to Eq. (\ref{pdftmin}) as well 
as new states to propagate; continue  
the propagation in the block $[\tau,2\tau)$. The procedure continues similarly 
for the rest of the propagation. The density of the system is then obtained 
as $n(\br,t)=\sum_{\alpha}\lket{\psi_{\alpha}(t)}\hat{n}(\br)\rket{\psi_{\alpha}(t)}$. 
The theorem discussed in section 2 also applies in this case. Therefore, the partition potential 
for this scheme is uniquely determined by the TD electronic density, up to an arbitrary constant. 
\begin{figure}[htb!]
\centering
\subfigure[$N_L$]{
\includegraphics[scale=0.25]{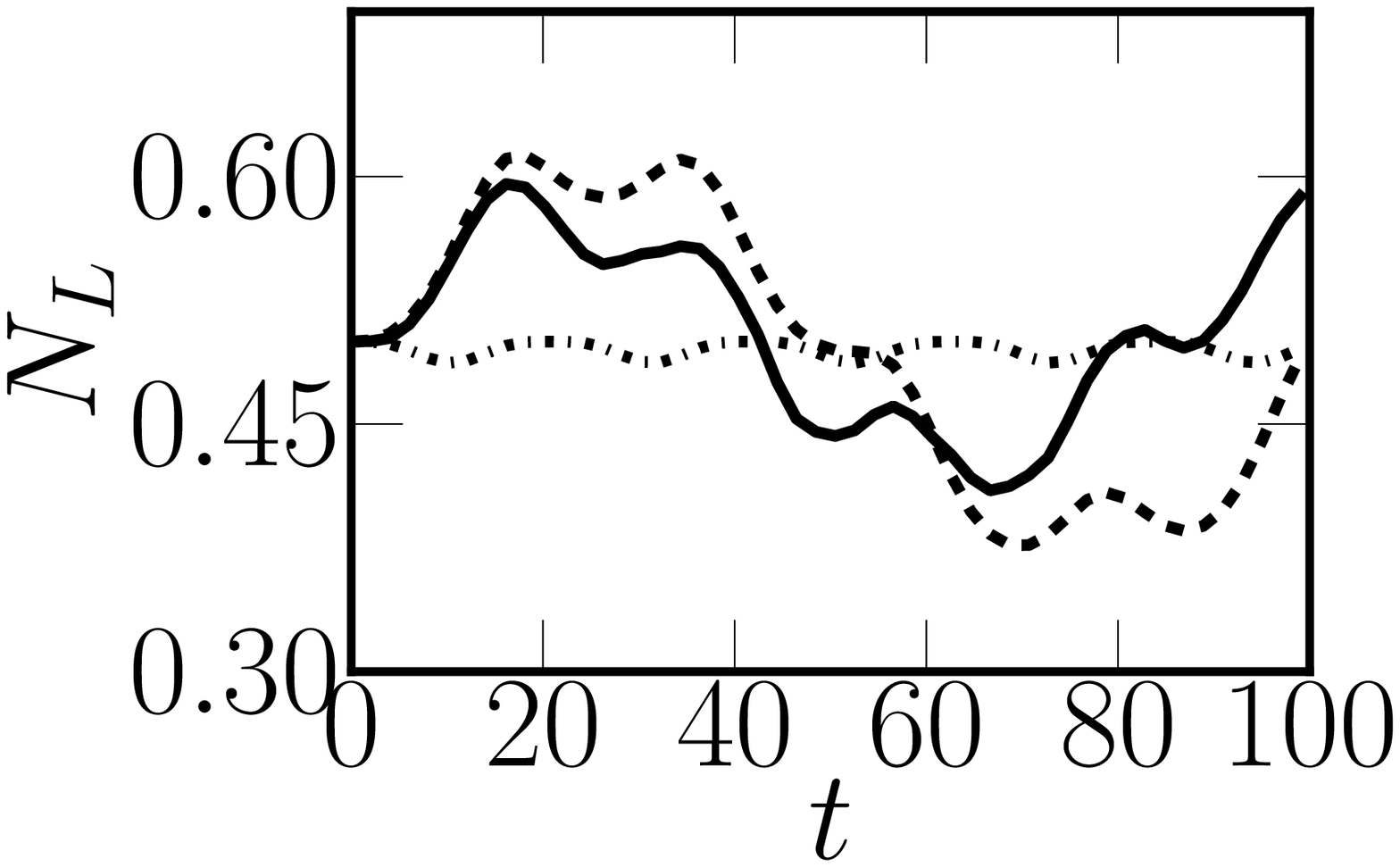}}\\
\subfigure[$v_{\mr{p}}(t=40)$]{
\includegraphics[scale=0.20]{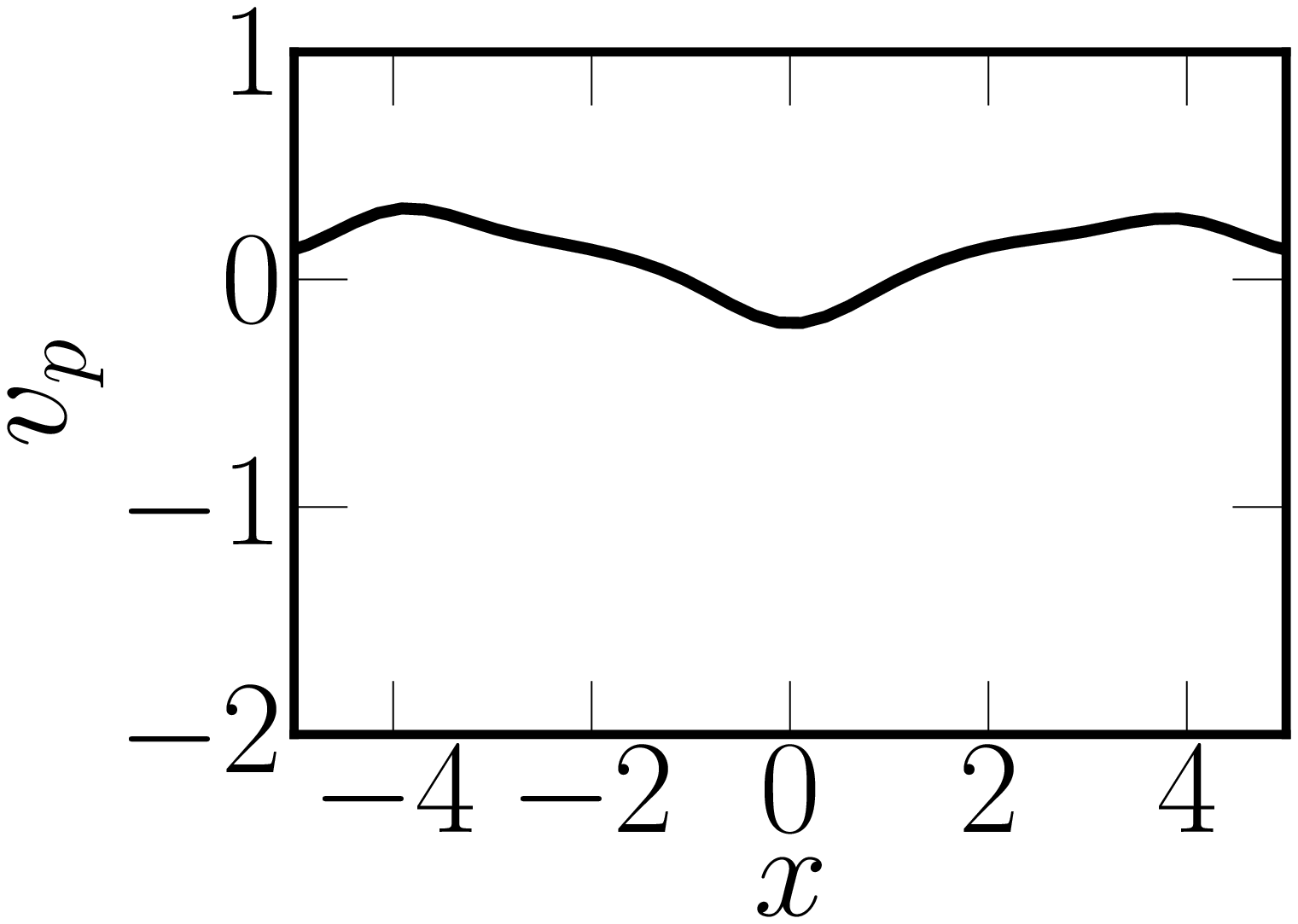}}
\subfigure[$n(t=40)$]{
\includegraphics[scale=0.20]{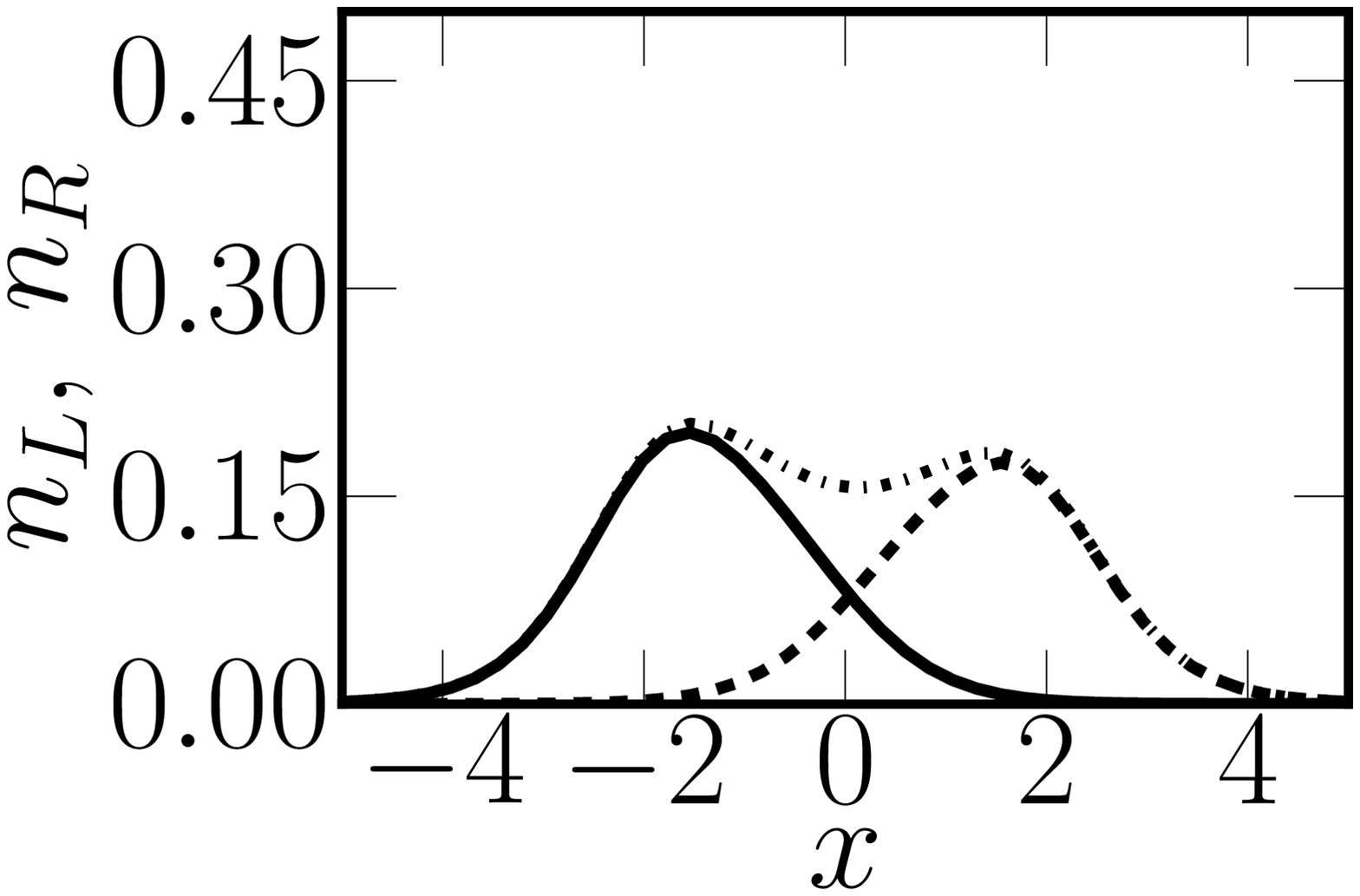}}
\subfigure[$v_{\mr{p}}(t=70)$]{
\includegraphics[scale=0.20]{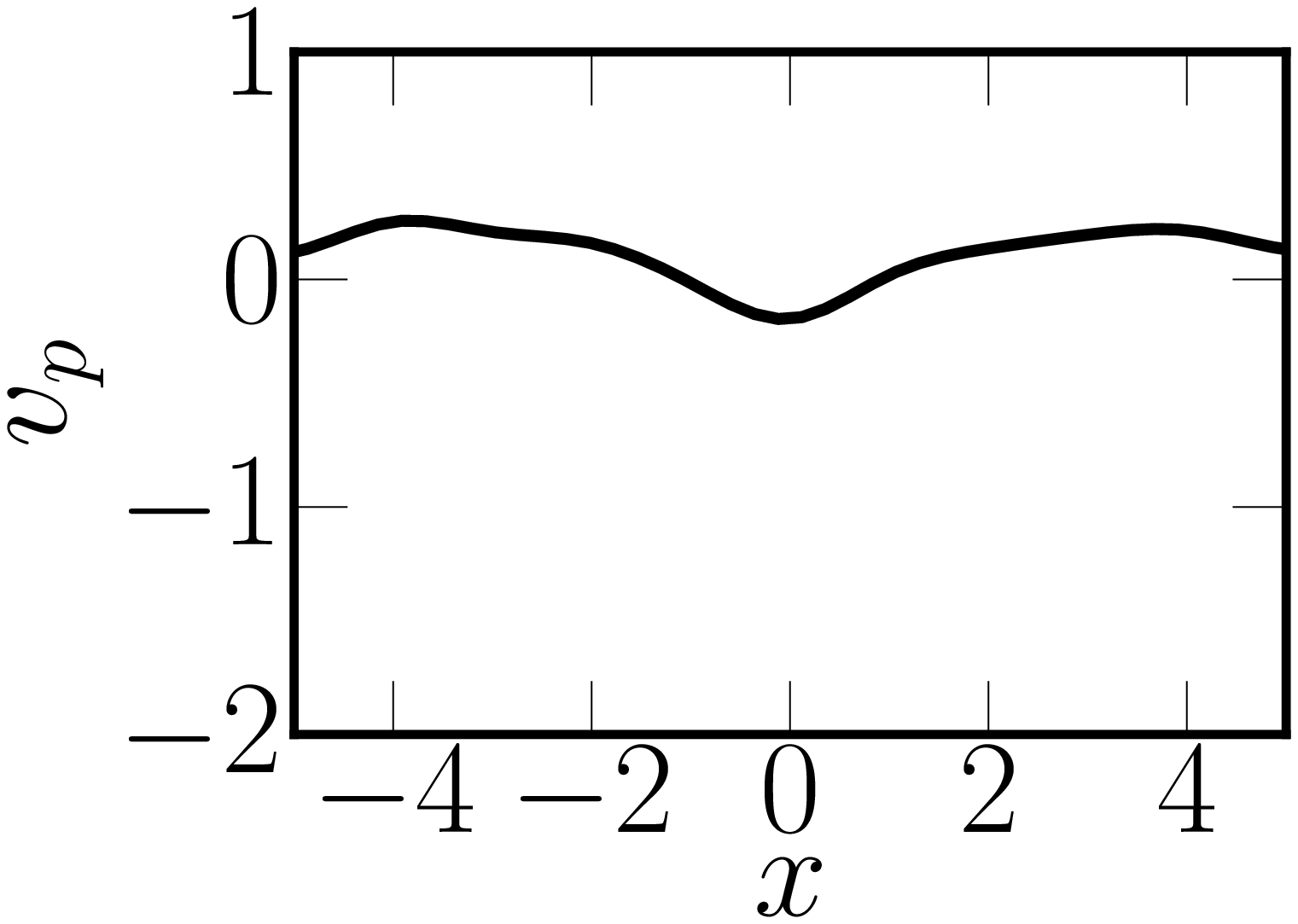}}
\subfigure[$n(t=70)$]{
\includegraphics[scale=0.20]{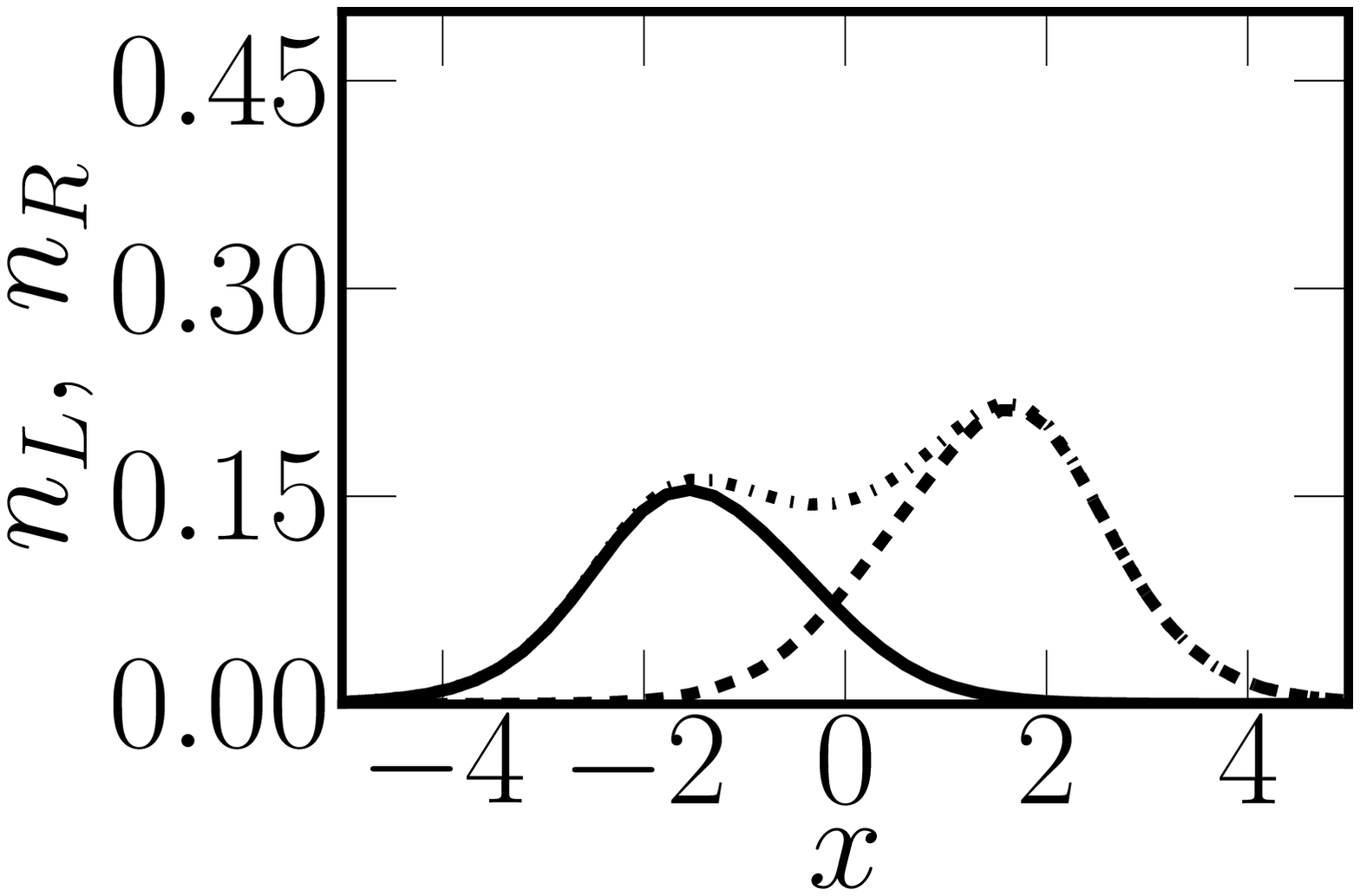}}
\caption{Evolution of the fragments with TD electron populations. In a) the solid line 
is the result from the inversion, the dashed line is the result from the two-state approximation, and 
the dashed-dotted line is obtained by omitting the gluing potential: $v_{\mr{G}}=0$. In c) and e), 
solid line: $n_{\mr{L}}$, dashed line: $n_{\mr{R}}$, dashed-dotted: $n$.}
\end{figure}

The partition potential is discontinuous at the relaxation nodes (points where 
$t$ is an integer multiple of $\tau$). Discontinuities in time can be eliminated by using an integral transformation 
that smooths the observable at the relaxation nodes. In practice, however, 
it is convenient to propagate the occupation numbers and gluing potential assuming that they are 
continuously differentiable functions of time. It can be shown, assuming that 
the dynamics of the occupation numbers is much slower than that of the partition potential, 
that the 1-1 map between the former and the density still holds. This follows 
from the scheme we have shown here because the electronic populations are fixed in the first 
block, allowing us to apply the Runge-Gross theorem in such block.

A density-functional approximation to the occupation numbers is needed
to apply the theory.
The dynamics of the occupation numbers can be investigated using master 
equations, where the rate coefficients are determined by Fermi's golden rule,
or transition elements that couple the fragments. Here, we illustrate 
a simple approach:
A trial wave function to investigate the evolution of the 
occupation numbers is $\rket{\eta(t)}=a_{\mr{L}}(t)\rket{\varphi_{\mr{L}}}
+a_{\mr{R}}(t)\rket{\varphi_{\mr{R}}}$, where $\rket{\varphi_{\alpha}}$ 
is the ground-state of the electron described only by $\hat{H}_{\alpha}^0$
(This hamiltonian in coordinate representation is $-1/2\partial^2_x+v_{\alpha}(x)$).
The dynamics of electron transfer is governed
by a two-component wave-function $\mb{a}=(a_{\mr{L}},a_{\mr{R}})^{\mr{T}}$.
We assume that the Hamiltonian coupling that relates the two fragments is of the form:
\ben
\hat{\mc{H}}(t)=\hat{H}_{\mr{f}}^0+\int \ud x~(v_{\mr{G}}(x,0)+v_{\mr{d}}(x,t))\hat{n}(x)
\een
where $\hat{H}_{\mr{f}}^0=\hat{H}_{\mr{L}}^0\oplus\hat{H}_{\mr{R}}^0$, is the uncoupled 
Hamiltonian; $\hat{H}_{\alpha}^0\rket{\varphi_{\beta}}=0$ if $\alpha\neq\beta$.
For the sake of the illustration, the gluing field is frozen,   
serving as a ``bridge'' for the charge to be transferred from one well into the other. 

From the evolution equation: $\ui \partial_t\rket{\eta(t)}=\hat{\mc{H}}(t)\rket{\eta(t)}$ we 
infer that the state vector, $\mb{a}$, satisfies:
\ben\ui\partial_t \mb{a}(t)=\mb{S}^{-1}(\bm{\epsilon}_0+\bm{\Delta}(t))\mb{a}(t)
\een
where $S_{\alpha\beta}=\int \ud x~\varphi_{\alpha}^*(x)\varphi_{\beta}(x)$, $\bm{\epsilon}_0=\mr{diag}(\epsilon_0,\epsilon_0)$, 
and 
\ben
\Delta_{\alpha\beta}(t)=\int \ud x~ \varphi_{\alpha}^*(x)(v_{\mr{G}}(x,0)+v_{\mr{d}}(x,t))\varphi_{\beta}(x)~.\een
The occupation numbers are obtained from the ``density'' of $\mb{a}$:
$N_{\alpha}(t)=|\mb{a}_{\alpha}|^2(t)+\mr{Re}(a^*_{\mr{L}}(t)a_{\mr{R}}(t)S_{\mr{LR}})$.
The last term arises from the overlap of the functions $\varphi_{\mr{L}}$ and $\varphi_{\mr{R}}$,
guaranteeing that $N_{\mr{L}}+N_{\mr{R}}=1$.

The example of the previous section, summarized in Fig. 1., illustrates how the partition potential
can be estimated even under 
a strong laser field. 
We now return to the example of section \ref{texample} and allow for variable occupations. The parameters for the propagation now
are $\tau=2$, $\Delta t=1$, $\omega=0.3$, $E_0=0.02$. 
Comparison of the exact time-dependency of the average number of electrons of the left fragment with 
the approximation described above is shown in Figure 3.a. The initial gluing potential 
used here is that shown in Fig. 1.a.
The two-state approximation works well at short times, and displays deviations after $t=20$.
Capturing the results of the two-state approximation would be quite challenging by fixing
the occupation numbers and finding the corresponding partition potential.
Improvements over the two-state approximation can proceed by either refining the gluing potential (going beyond 
the frozen approximation) or increasing the number of states 
considered to couple the fragments. The first alternative has the advantage that the 
equations can be solved very fast. Nonetheless, for functional development, the gluing potential is also a determinant 
factor for the evolution of the shape of the electronic fragment-density ($\int n_{\alpha}/N_{\alpha}$).

Figure 3.b shows a snapshot of the ``exact'' partition potential at $t=40$. In contrast with 
the results of section 3.2, the partition potential is now well localized (Figures 3.d and 3.f).
This suggests that the standard methods of ground-state PDFT can be used to estimate 
the partition potential through the use of the adiabatic approximation.
The fragment densities also remain localized (Figure 3.c, 3.e, and 3.g). Qualitatively, 
the partition potential accounts for the shape of the electronic densities 
of the fragments, while the occupation numbers are responsible for their height.

\section{Concluding Remarks}
We formulated a TDDFT for treating molecules as composed of smaller composite 
units. The successful application of this formulation requires 
approximations to the partition potential and the occupation numbers. This can 
be accomplished by a proper approximation to the Hamiltonians $\{\hat{H}_{{\mr{c}},{\alpha}}(t)\}$, 
or the auxiliary evolution equations of the electron populations in the fragments.
The error analysis was also presented. It leads to a simple form of estimating the 
errors in the potentials. The partition potential, obtained 
by numerical inversion, can be uncertain in spatio-temporal regions where the density is small. 
However, as time increases, the error 
can propagate from the boundary areas into regions were the density is high. 

\section{Acknowledgments}
We acknowledge support from the National Science Foundation CAREER program under grant No.CHE-1149968.
AW also acknowledges support from the Alfred P. Sloan Foundation and the Camille Dreyfus Teacher-Scholar
Awards Programs.

\bibliography{refs_ftdft}

\begin{thebibliography}{20}
\providecommand{\natexlab}[1]{#1}
\providecommand{\url}[1]{\texttt{#1}}
\expandafter\ifx\csname urlstyle\endcsname\relax
  \providecommand{\doi}[1]{doi: #1}\else
  \providecommand{\doi}{doi: \begingroup \urlstyle{rm}\Url}\fi

\bibitem[Gross and Maitra(2012)]{GM12}
Eberhard K.~U. Gross and Neepa~T. Maitra.
\newblock {Introduction to TDDFT Fundamentals of Time-Dependent Density
  Functional Theory}.
\newblock volume 837 of \emph{{Lecture Notes in Physics}}, chapter~4, pages
  53--99. Springer Berlin / Heidelberg, Berlin, Heidelberg, 2012.
\newblock ISBN 978-3-642-23517-7.

\bibitem[Runge and Gross(1984)]{RG84}
Erich Runge and E.~K.~U. Gross.
\newblock {Density-Functional Theory for Time-Dependent Systems}.
\newblock \emph{Phys. Rev. Lett.}, 52\penalty0 (12):\penalty0 997, March 1984.

\bibitem[Peuckert(1978)]{P78}
Volker Peuckert.
\newblock A new approximation method for electron systems.
\newblock \emph{J. Phys. C: Solid State Physics}, 11\penalty0 (24):\penalty0
  4945, 1978.

\bibitem[Zangwill and Soven(1980)]{ZS80}
A~Zangwill and Paul Soven.
\newblock Resonant photoemission in barium and cerium.
\newblock \emph{Phys. Rev. Lett.}, 45\penalty0 (3):\penalty0 204, 1980.

\bibitem[Casida and Weso{\l}owski(2004)]{CW04}
Mark~E. Casida and Tomasz~A. Weso{\l}owski.
\newblock {Generalization of the Kohn--Sham equations with constrained electron
  density formalism and its time-dependent response theory formulation}.
\newblock \emph{Int. J. Quantum Chem.}, 96\penalty0 (6):\penalty0 577--588,
  2004.

\bibitem[Mahito et~al.(2007)Mahito, Fedorov, and Kitaura]{CFK07}
Chiba Mahito, Dmitri~G. Fedorov, and Kazuo Kitaura.
\newblock {Time-dependent density functional theory based upon the fragment
  molecular orbital method}.
\newblock \emph{J. Chem. Phys.}, 127:\penalty0 104108, 2007.

\bibitem[Neugebauer(2010)]{N10}
Johannes Neugebauer.
\newblock {Chromophore-specific theoretical spectroscopy: From subsystem
  density functional theory to mode-specific vibrational spectroscopy}.
\newblock \emph{Phys. Rep.}, 489\penalty0 (1-3):\penalty0 1--87, April 2010.
\newblock ISSN 03701573.

\bibitem[Pavanello(2013)]{P13}
Michele Pavanello.
\newblock On the subsystem formulation of linear-response time-dependent dft.
\newblock \emph{J. Chem. Phys.}, 138\penalty0 (20):\penalty0 204118, 2013.

\bibitem[Neugebauer(2007)]{N07}
Johannes Neugebauer.
\newblock {Couplings between electronic transitions in a subsystem formulation
  of time-dependent density functional theory}.
\newblock \emph{J. Chem. Phys.}, 126\penalty0 (13):\penalty0 134116, 2007.

\bibitem[Severo Pereira~Gomes and Jacob(2012)]{PJ12}
Andre Severo Pereira~Gomes and Christoph~R. Jacob.
\newblock {Quantum-chemical embedding methods for treating local electronic
  excitations in complex chemical systems}.
\newblock \emph{Annu. Rep. Prog. Chem., Sect. C: Phys. Chem.}, 108\penalty0
  (1):\penalty0 222--277, 2012.

\bibitem[Mosquera et~al.(2013)Mosquera, Jensen, and Wasserman]{MJW13}
Mart{\'\i}n~A Mosquera, Daniel Jensen, and Adam Wasserman.
\newblock Fragment-based time-dependent density functional theory.
\newblock \emph{Phys. Rev. Lett.}, 111\penalty0 (2):\penalty0 023001, 2013.

\bibitem[Mosquera and Wasserman(2014)]{MW14}
Mart{\'\i}n~A Mosquera and Adam Wasserman.
\newblock Current density partitioning in time-dependent current density
  functional theory.
\newblock \emph{J. Chem. Phys.}, 140\penalty0 (18):\penalty0 18A525, 2014.

\bibitem[Huang et~al.(2014)Huang, Libisch, Peng, and Carter]{HLPC14}
Chen Huang, Florian Libisch, Qing Peng, and Emily~A Carter.
\newblock Time-dependent potential-functional embedding theory.
\newblock \emph{J. Chem. Phys.}, 140\penalty0 (12):\penalty0 124113, 2014.

\bibitem[Elliott et~al.(2010)Elliott, Burke, Cohen, and Wasserman]{EBCW10}
Peter Elliott, Kieron Burke, Morrel~H. Cohen, and Adam Wasserman.
\newblock {Partition density-functional theory}.
\newblock \emph{Phys. Rev. A}, 82\penalty0 (2):\penalty0 024501, August 2010.

\bibitem[Mosquera and Wasserman(2013)]{MW12}
Mart\'{\i}n~A. Mosquera and Adam Wasserman.
\newblock {Partition density functional theory and its extension to the
  spin-polarized case}.
\newblock \emph{Mol. Phys.}, 111\penalty0 (4):\penalty0 505--515, September
  2013.

\bibitem[Cohen and Wasserman(2006)]{CW06}
Morrel~H. Cohen and Adam Wasserman.
\newblock {On Hardness and Electronegativity Equalization in Chemical
  Reactivity Theory}.
\newblock \emph{J. Stat. Phys.}, 125\penalty0 (5):\penalty0 1121--1139,
  December 2006.
\newblock ISSN 0022-4715.

\bibitem[Castro et~al.(2012)Castro, Werschnik, and Gross]{CWG12}
Alberto Castro, Jan Werschnik, and Eberhard~KU Gross.
\newblock Controlling the dynamics of many-electron systems from first
  principles: a combination of optimal control and time-dependent
  density-functional theory.
\newblock \emph{Phys. Rev. Lett.}, 109\penalty0 (15):\penalty0 153603, 2012.

\bibitem[Ruggenthaler and van Leeuwen(2011)]{RvL11}
Michael Ruggenthaler and Robert van Leeuwen.
\newblock Global fixed-point proof of time-dependent density-functional theory.
\newblock \emph{Europhys. Lett.}, 95\penalty0 (1):\penalty0 13001, 2011.

\bibitem[Seber and Wild(1989)]{SW89}
GAF Seber and CJ~Wild.
\newblock \emph{Nonlinear Regression}.
\newblock Wiley, New York, 1989.

\bibitem[Roncero et~al.(2015)Roncero, Aguado, Batista-Romero, Bernal-Uruchurtu,
  and Hern{\'a}ndez-Lamoneda]{RABB15}
Octavio Roncero, Alfredo Aguado, Fidel~A Batista-Romero, Margarita~I
  Bernal-Uruchurtu, and Ram{\'o}n Hern{\'a}ndez-Lamoneda.
\newblock Density-difference-driven optimized embedding potential method to
  study the spectroscopy of br2 in water clusters.
\newblock \emph{J. Chem. Phys.}, 11\penalty0 (3):\penalty0 1155--1164, 2015.

\end{thebibliography}

\end{document}